\documentclass[aps,prd, onecolumn,amssymb]{revtex4}
\usepackage{graphicx,bm}
\usepackage{amsmath}
\usepackage{amsfonts}
\begin{document}

\newcommand{\be}{\begin{equation}}
\newcommand{\ee}{\end{equation}}
\newcommand{\bea}{\begin{eqnarray}}
\newcommand{\eea}{\end{eqnarray}}
\newcommand{\nn}{\nonumber \\}
\newcommand{\e}{\mathrm{e}}

\title{Effective Dark Energy Models and Dark Energy Models with Bounce in frames of $F(T)$ Gravity}
\author{Artyom~V.~Astashenok}
\affiliation{I. Kant Baltic Federal University, Institute of Physics and Technology, 236041, 14, Nevsky st., Kaliningrad, Russia}

\begin{abstract}

Various cosmological models in frames of $F(T)$ gravity are considered. The general scheme of constructing effective dark energy models with various evolution is presented. It is showed that these models in principle are compatible with $\Lambda$CDM model. The dynamics of universe governed by $F(T)$ gravity can mimics $\Lambda$CDM evolution in past but declines from it in a future. We also construct some dark energy models with the ``real'' (non-effective) equation-of-state parameter $w$ such that $w\leq-1$. It is showed that in $F(T)$ gravity the Universe filled phantom field not necessarily ends its existence in singularity. There are two possible mechanisms permitting the final singularity. Firstly due to the nonlinear dependence between energy density and $H^{2}$ ($H$ is the Hubble parameter) the universe can expands not so fast as in the general relativity and in fact Little Rip regime take place instead Big Rip. We also considered the models with possible bounce in future. In these models the universe expansion can mimics the dynamics with future singularity but due to bounce in future universe begin contracts.
\end{abstract}

\maketitle

\section{Introduction}

The problem of initial singularity in the cosmological models based on
General Relativity (GR) is one of the puzzles of modern physics. For resolution of this problem many approaches are offered such as null-energy-condition violating quantum fluctuations  \cite{dutta1, dutta2}, quantum gravity effects, or effective field theory techniques.

The initial singularity can be avoided in frames of non-singular bouncing cosmological models \cite{Mukhanov:1991zn}. The key feature of such models is modification of standard Einstein-Hilbert action.  The Pre-Big-Bang \cite{Veneziano:1991ek} and the Ekpyrotic
\cite{Khoury:2001wf} models, gravitational actions with higher order
corrections \cite{Brustein:1997cv}, braneworld scenarios
\cite{Kehagias:1999vr}, loop quantum cosmology
\cite{Bojowald:2001xe} are investigated in detail. {{For a review on models of modifying higher
derivative gravity, which solve cosmic singularities in an efficient way,
we refer to Ref. \cite{Nojiri:2010wj}.}}

Above mentioned theories also attract attention for explaining of accelerated expansion of the universe \cite{Riess}, \cite{Perlmutter}. In frames of standard GR cosmology for explaining current acceleration one need to introduce the dark energy phenomena. The dark energy has negative pressure (for review see \cite{Dark-1,Dark-2,Dark-3,Dark-4,Dark-5,Dark-6,Cai:2009zp,Cai:2006dm}). In frames of $\Lambda$CDM model in which the dark energy is simply cosmological constant ($p=-\rho$) the observational data require that dark matter and baryonic matter 27.4\% of universe energy \cite{Kowalski}.
The equation-of-state parameter
$w_{D}$ for dark energy is negative:
\be
w_{D}=p_{D}/\rho_{D}<0\, ,
\ee
where $\rho_{D}$ is the dark energy density and $p_{D}$ is the pressure. If $w<-1$ the
violation of all four energy conditions  occurs. The corresponding phantom
field, which is instable as quantum field theory \cite{Carrol} but could be
stable in classical cosmology  may be naturally described by the scalar field
with the negative kinetic term.

Phantom cosmology may lead to so-called Big Rip singularity
\cite{Starobinsky,Caldwell,Frampton,Nesseris,Diaz,Nojiri}. There are several main ways to avoid the Big Rip singularity:

(i) for some scalar potentials one can consider phantom acceleration as transient phenomenon.

(ii) probably the accounting of quantum effects may delay/stop  the singularity occurrence \cite{Nojiri-2}.

(iii) in theories of modified gravity one can construct cosmological models which can mimics phantom acceleration but free from singularities (for review,
see \cite{review},\cite{Bamba-3},\cite{Bamba-4}).

Besides the Big Rip singularity another finite-time singularities
may occur (for classification see \cite{Nojiri-3}.

Theoretical description of dark energy often use the formalism of equation-of-state. For pressure of dark
energy one can choose the general expression
\be \label{EoS}
p=-\rho-g(\rho)\, ,
\ee
where $g(\rho)$ is function of energy density. The case $g(\rho)>0$ corresponds
to $w<-1$.

One note that in frames of EoS formalism one can construct phantom energy models with $w<-1$ but without Big Rip singularity. There are two possibilities: $w$ asymptotically tends to $-1$ and energy density increases with time (Little Rip, see \cite{Frampton-2,Frampton-3,LR,Frampton-4,Liu:2012iba}) or remains constant \cite{Yurov-4,Nojiri-3,barrow}.
The key moment is that if $w$ approaches $-1$ sufficiently rapidly,
then it is possible to have a model in which the time required for singularity
is infinite, i.e., singularity effectively does not occur.

The aim of this article is constructing the phantom models in frames of $f(T)$ gravity. This modification of gravity is more simple than $f(R)$ gravity which lead to fourth-order field equations. The reconstruction of different cosmological models in frames of $F(T)$ gravity can be performed \cite{Myrzakulov:2010vz}.

In frames of $F(T)$ theory the null energy condition could be effectively violated and therefore one can obtain the cosmological solutions with acceleration but without dark energy (``effective dark energy model'') and models with bounce in the early universe or in the future. The initial or future singularities can be avoided in $F(T)$ gravity.

This paper is organized as follows. In the next section we briefly describe the key equations of $f(T)$ cosmology. Further the general method of constructing ``effective dark energy'' models (i.e. models in which universe expands with acceleration containing baryon and dark matter only) is presented. We also consider the phantom energy in frames of $F(T)$ gravity with various EoS and investigate the possibilities of avoiding of big rip singularity. The last section is devoted to conclusions.

\section{$f(T)$ gravity and cosmology}

We consider a flat Friedmann-Robertson-Walker spacetime with metric
\begin{equation}
ds^2= dt^2-a^2(t)\,\delta_{ij} dx^i dx^j,
\end{equation}
where $a(t)$ is the scale factor.
The action in f(T) gravity can be written as
\begin{equation}  \label{action}
S = \frac{1}{2}\int d^4x e \left[T+f(T)+L_m\right],
\end{equation}
where $T$ is the torsion scalar $T$. For $f(T)=0$ the action (\ref{action}) is completely equivalent to GR.

The derivation of the equations is described in detail in \cite{main}. We follow to this work. In this approach dynamical variables are vierbein fields ${\mathbf{e}_A(x^\mu)}$. One can define with these fields an orthonormal basis for the tangent space at each point, i.e. $\mathbf{e}_A\cdot%
\mathbf{e}_B=\eta_{AB}$, where $\eta_{AB}=\mbox{diag}(1,-1,-1,-1)$. The components $e_A^\mu$ of vector $\mathbf{e}_A$ i are $\mathbf{e}_A=e^\mu_A\partial_\mu $.

For metric tensor we have the following relation
\begin{equation}  \label{metrdef}
g_{\mu\nu}(x)=\eta_{AB}\, e^A_\mu (x)\, e^B_\nu (x).
\end{equation}
In $F(T)$ gravity the curvatureless Weitzenb\"{o}ck connection is used instead
the torsionless connection in GR.

The ``teleparallel Lagrangian'' is simply torsion scalar, i.e.
\begin{equation}  \label{telelag}
T\equiv S_\rho^{\:\:\:\mu\nu}\:T^\rho_{\:\:\:\mu\nu},
\end{equation}
where
\begin{equation}  \label{Stensor}
S_\rho^{\:\:\:\mu\nu}=\frac{1}{2}\Big(K^{\mu\nu}_{\:\:\:\:\rho}
+\delta^\mu_\rho
\:T^{\alpha\nu}_{\:\:\:\:\alpha}-\delta^\nu_\rho\:
T^{\alpha\mu}_{\:\:\:\:\alpha}\Big),
\end{equation}
and
\begin{equation}  \label{cotorsion}
K^{\mu\nu}_{\:\:\:\:\rho}=-\frac{1}{2}\Big(T^{\mu\nu}_{
\:\:\:\:\rho}
-T^{\nu\mu}_{\:\:\:\:\rho}-T_{\rho}^{\:\:\:\:\mu\nu}\Big).
\end{equation}
is difference between the Weitzenb%
\"{o}ck and Levi-Civita connections.
The torsion tensor is defined as
\begin{equation}  \label{torsion2}
{T}^\lambda_{\:\mu\nu}=\overset{\mathbf{w}}{\Gamma}^\lambda_{
\nu\mu}-%
\overset{\mathbf{w}}{\Gamma}^\lambda_{\mu\nu}
=e^\lambda_A\:(\partial_\mu
e^A_\nu-\partial_\nu e^A_\mu).
\end{equation}

The equations of motion can be obtained by varying the action
(\ref{action}) with respect to the vierbein:
\begin{widetext}
\begin{eqnarray}\label{eom}
e^{-1}\partial_{\mu}(eS_{A}{}^{\mu\nu})[1+f_{,T}]
-e_{A}^{\lambda}T^{\rho}{}_{\mu\lambda}S_{\rho}{}^{\nu\mu} +
S_{A}{}^{\mu\nu}\partial_{\mu}({T})f_{,TT}-\frac{1}{4}e_{A}^{\nu
}[T+f({T})]
= \frac{1}{2}e_{A}^{\rho}\overset {\mathbf{em}}T_{\rho}{}^{\nu},
\end{eqnarray}
\end{widetext}
where $f_{,T}$ and $f_{,TT}$  are the
first and second derivatives of the function $f(T)$ with respect
to $T$, and
the mixed indices are used as in $S_A{}^{\mu\nu} =
e_A^{\rho}S_{\rho}{}^{\mu\nu}$. The energy-momentum tensor is
$\overset{\mathbf{em}}{T%
}_{\rho}{}^{\nu}$.

For universe filled ``perfect fluid" with $\overset{\mathbf{em}}{T}_{\mu \nu
}=\mbox{diag}(\rho,-p,-p,-p)$ where $\rho$ and $p$ are the
energy density and pressure of the matter content, one can see that (\ref{eom}) lead to the
Friedmann equations
\begin{eqnarray}
&&H^{2}=\frac{\rho}{3}-\frac{f({T})}{6}-2f_{,T}H^{2}
\label{fried1} \\
&&\dot{H}=-\frac{\rho+p}{2(1+f_{,T}-12H^2f_{,TT})%
},  \label{fried2}
\end{eqnarray}%
where $H$ is the Hubble parameter $H\equiv \dot{a}/a$, where a dot denotes a derivative with respect to coordinate
time $t$.
One should to mention the useful relation between torsion scalar and Hubble parameter
\begin{equation}
T=-6H^{2},  \label{TH2}
\end{equation}%
which can be easily derived from evaluation of (\ref{telelag}). As one can expect from (\ref{fried1}) and (\ref{fried2}) the equation of continuity follows
\be\label{fried3}
\dot{\rho}+3H(\rho+p)=0.
\ee
For given function $f(T)$ one can consider the parameter $H$ as function of dark energy density, $H=H(\rho)$.

\section{Dark energy cosmology in $F(T)$ gravity}

\textbf{Effective dark energy model in frames of $F(T)$ gravity.}

Let's briefly review the possibilities of constructing effective dark energy models in $F(T)$ gravity. There are alternative to $\Lambda$CDM model \cite{Wu:2010xk}, \cite{Chen:2010va,Dent:2011zz}, \cite{Li2011}. We assume that universe contains only cold dark matter with $p=0$, neglecting radiation. Then from the (\ref{fried2}) one can obtain link between time and Hubble parameter
\be\label{TFT}
t=-\frac{1}{3}\int\frac{dx}{x^{1/2}}\frac{d}{dx}\ln(6x+f(x)-2f'(x)x),
\ee
where $x=H^{2}$ and prime designates the derivative on $x$. If universe expands so that $\dot{H}>0$, i.e. $3-f'(x)-xf''(x)<0$ the matter density
\be
\rho=3x+f(x)/2-x f'(x)
\ee
decreases with growing $H$ ($\rho'(x)<0$) as might have been expected. Integrating by part we have
\be\label{TFT-2}
t=-\frac{1}{3}\frac{\ln(6x+f(x)-2xf'(x))}{x^{1/2}}\mid^{x}_{x_{0}}-\frac{1}{6}\int_{x_0}^{x}\frac{dx}{x^{3/2}}\ln(6x+f(x)-2xf'(x)).
\ee

From this relation one can see that there are following possibilities defined by behavior of first term:

(i) the first term converges at $x\rightarrow\infty$ (the second integral in this case also converges) and finite-time singularity occurs.

(ii) the first term diverges at $x\rightarrow\infty$. We have the little rip: universe expands with increasing acceleration but singularity effectively does not occur.

(iii) $t\rightarrow\infty$ $x\rightarrow x_{f}$, i.e. the rate of expansion tends to constant value in future. Asymptotic de Sitter regime realizes.

Choosing the dependence $6x+f(x)-2xf'(x)$ one can define the function $f(x)$ and then matter density as function of $x$. Further taking into account that
$$
\rho=\rho_{0}/a^{3}
$$
one can derive the time evolution of scale factor. For comparison with observational data (for example, modulus vs redshift relation for SNe Ia) one need the dependence $H(z)/H_{0}$ from redshift $z$.

We take example. Let's $6x+f(x)-2xf'(x)=3(x-x_{\lambda})\exp(-\beta x^{\gamma})$, $\beta>0$ and $x>x_{\lambda}$. Then we have the following relation for determination $H$ as function of redshift:
\be
\frac{\rho_{m0}}{(1+z)^{3}}=3(H^{2}-H^{2}_{\lambda})\exp(-\beta H^{2\gamma}).
\ee
For $\beta<<H^{-2\gamma}$ we have that our equation gives that
$$
H^{2}\approx H^{2}_{\lambda}+\frac{\rho_{m0}}{(1+z)^{3}}.
$$
This relation corresponds to $\Lambda$CDM cosmology.
From (\ref{TFT-2}) one see that for $\gamma\leq 1/2$ finite-time singularity occurs. The little rip regime realizes if $\gamma>1/2$.
We also present the form of the function $f(x)$ in a case when $\gamma=1$:
\be
f(x)=6x-\left(\frac{3}{2}\sqrt{\frac{\pi}{\beta}}\mbox{erf}(\sqrt{\beta x})+3x_{\lambda}\left[\sqrt{\pi\beta}\mbox{erf}(\sqrt{\beta x})+x^{-1/2}\exp(-\beta x)\right]\right)x^{1/2},
\ee
where $\mbox{erf}$ is error function.

\textbf{Phantom energy models in frames of F(T) gravity.}

Now let's consider the features of cosmological models of phantom energy in $F(T)$ gravity. We will examine the future evolution of our universe from the point at which
the pressure and density are dominated by the dark energy. Using (\ref{EoS}) from (\ref{fried3}), one can obtain the following
link between time and $f(\rho)$:
\be
\label{trho}
t = \frac{1}{3}\int^{\rho}_{\rho_{0}} \frac{d \rho}{H(\rho)f(\rho)}.
\ee
The various possibilities of universe evolution in frames of usual FRLW cosmology are described in detail \cite{Astashenok}. Here we note the some features which can take place in $F(T)$ background.

\textbf{1. Avoiding the Big Rip singularity due to ``softening'' of EoS}

Let's consider the case $f(T)=\alpha(-T/6)^{\beta}$, where $\alpha$ and $\beta$ are some constants. Using the relation $T=-6H^{2}$ one can rewrite the equation (\ref{fried1}) in the following form:
\be\label{frA}
((-2\beta+1)\alpha(H^{2})^{\beta-1}+1)H^{2}=\frac{\rho}{3}.
\ee
Assuming $\beta>1$, $\alpha<0$ one can conclude that if $|\alpha|<<(2\beta-1)^{-1}H_{0}^{2(1-\beta)}$ ($H_{0}$ is the current Hubble parameter) than the Eq. (\ref{frA}) differs from Friedmann equation negligibly. But from (\ref{fried2}) follows that Hubble parameter grows with time and the first term in brackets of (\ref{frA}) on late times becomes the dominant. We have the following asymptotic for Hubble parameter:
$$
H\sim \rho^{1/2\beta}, \quad t>>0.
$$
Considering the case $g\sim \rho^{\gamma}$, $\gamma>0$ one can see that in a case of GR cosmology the big singularity occurs if $1/2<\gamma\leq1$. But in a case of $F(T)$ gravity the condition of non-singular evolution is milder: $\gamma\leq1-1/2\beta$. Therefore if $1/2<\gamma\leq1-1/2\beta$ we have little rip dynamics in a future instead big rip singularity as in general relativity. Therefore the ``effective'' EoS on late times corresponds to initial with replacing $\gamma\rightarrow\gamma-1/2+1/2\beta<\gamma$.

For example one consider the case $\beta=2$. From Eq.(\ref{frA}) it is easy to obtain for Hubble parameter:
\be
H^{2}=\frac{1}{6\alpha}\left(1-\sqrt{1-4\alpha\rho}\right).
\ee
For $\rho\ll|\alpha|^{-1}$ it follows that $H^{2}\approx\rho/3$. If $\rho>>|\alpha|^{-1}$ we have $H^{2}\approx (1/3)(\rho/\alpha)^{1/2}$.

\textbf{2. Models with bounce.}

Let's consider the Eq. (\ref{fried1}) as differential equation on function $f(T)$. Using relation (\ref{TH2}) one can rewrite this equation as
\be \label{EQFT}
f'(x)-\frac{f(x)}{2x}=-3+\frac{\rho(x)}{x}.
\ee
We consider the energy density as function of $H^{2}$ in this equation. For given EoS one can derive the dependence $\rho=\rho(a)$ and choosing scale factor as function of time we obtain the dependence $\rho=\rho(H^2)$ (in general case only in parametric form of course).
The solution of (\ref{EQFT}) is
\be
f(x)=Cx^{1/2}-6x-x^{1/2}\int \frac{\rho(x)dx}{x^{3/2}},
\ee
where $C$ is integration constant. Using the continuity equation (\ref{fried3}) we have for $f$ as function of time
\be \label{EQFT-2}
f(t)=Cx^{1/2}(t)-6x(t)+2\rho(t)-6x^{1/2}(t)\int g(\rho(t))dt.
\ee
The generalization of (\ref{EQFT}) on a case of universe filled n-component fluid is simple. One note that $g(\rho)=-(\rho+p)$ and therefore the Eq. (\ref{EQFT}) turns into
\be \label{EQFT-1}
f(t)=Cx^{1/2}(t)-6x(t)+2\sum_{i=1}^{n}\rho_{i}(t)+6x^{1/2}(t)\sum_{i=1}^{n}\int (\rho_{i}(t)+p_{i}(t))dt.
\ee
Further for simplicity we put $C=1$.

One can simply construct the cosmological models without singularities (past or future).

\textbf{$\Lambda$CDM model with bounce in past}. In \cite{main} the simplest version of an $f(T)$ matter bounce is considered. One can construct little more complex model following to methodology \cite{main}. Let's consider the universe filled cold dark matter and vacuum energy with density $\Lambda$. In this case the evolution of scale factor in GR cosmology can be written in form
\begin{equation}\label{LCDMA}
a(t)=a_{0}\sinh^{2/3}(t/T),
\end{equation}
where $T=2\sqrt{1/3\Lambda}$. Our moment of time is $t_{0}=T\mbox{arcth}\Omega_{\Lambda}^{1/2}$, $\Omega_{\Lambda}=\Lambda/(\rho_{m0}+\Lambda)$. The moment $t=0$ corresponds to Big Bang singularity.
In $f(T)$ gravity one can consider the such evolution of scale factor which close to (\ref{LCDMA}) at $t>>\tau$, where $\tau$ is the moment of time such that $\tau<<T$. For example let's choose
\begin{equation}
a_{b}(t)=a_{0}\sinh^{2/3}\left(\frac{(t^{2}+\tau^{2})^{1/2}}{T}\right).
\end{equation}
This dynamics mimics $\Lambda$CDM model: for $\ll \tau<t\ll T$ we have $a_{b}\sim t^{2/3}$ and for $t>>T$ $a_{b}(t)\sim \exp (\sqrt{\Lambda/3} t) $. At the moment $t=0$ the bounce occurs instead Big Bang singularity. The time variable can be varies in interval $-\infty<t<\infty$.

For function $f(t)$ the following relation take place
\begin{equation}
f(t)=\frac{8}{3T^{2}}\frac{\tau^{2}}{\tau^{2}+t^{2}}\coth^{2}\left(\frac{(t^{2}+\tau^{2})^{1/2}}{T}\right)
+\frac{16t}{3T^{3}(t^2+\tau^{2})^{1/2}}\coth\left(\frac{(t^{2}+\tau^{2})^{1/2}}{T}\right)\int\sinh^{-2}\left(\frac{(t^{2}+\tau^{2})^{1/2}}{T}\right)dt.
\end{equation}
Introducing dimensionless time $\tilde{t}=t/T$ one can rewrite this relation in the form
\begin{equation}
f(\tilde{t})=2\Lambda\frac{\sigma^{2}}{\sigma^{2}+\tilde{t}^{2}}\coth^{2}(\sigma^{2}+\tilde{t}^{2})^{1/2}
+4\Lambda\frac{\tilde{t}\coth(\sigma^{2}+\tilde{t}^{2})^{1/2}}{(\tilde{t}^2+\sigma^{2})^{1/2}}\int\sinh^{-2}(\sigma^{2}+\tilde{t}^{2})^{1/2}d\tilde{t}.
\end{equation}
For $T$ as function of $\tilde{t}$ one have
\begin{equation}
T=-2\Lambda\frac{\tilde{t}^{2}\coth^{2}(\sigma^{2}+\tilde{t}^{2})^{1/2}}{(\tilde{t}^2+\sigma^{2})}.
\end{equation}

In order to present function $f(T)$ more clearly,
in Fig. 1 we depict $f(T)$ for some values of $\sigma$. We also derived the evolution of the $f(T)$ and the Hubble parameter as functions of the
dimensionless cosmic time on Figs. 2 and 3 respectively. Our results as expected are close to \cite{main}.
\begin{figure}
\includegraphics{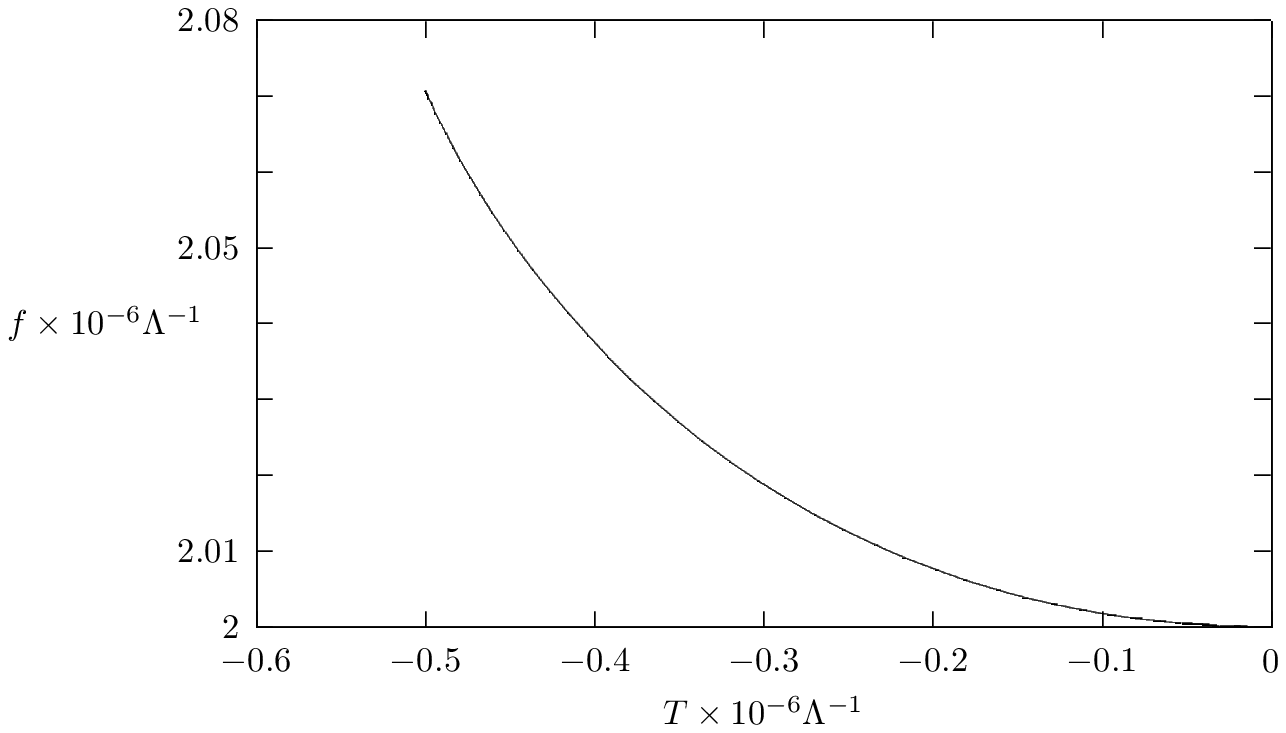}\\
\caption{The form of f as a function of the torsion scalar $T$
in $\Lambda$CDM bounce cosmology.}
\end{figure}

\begin{figure}
\includegraphics{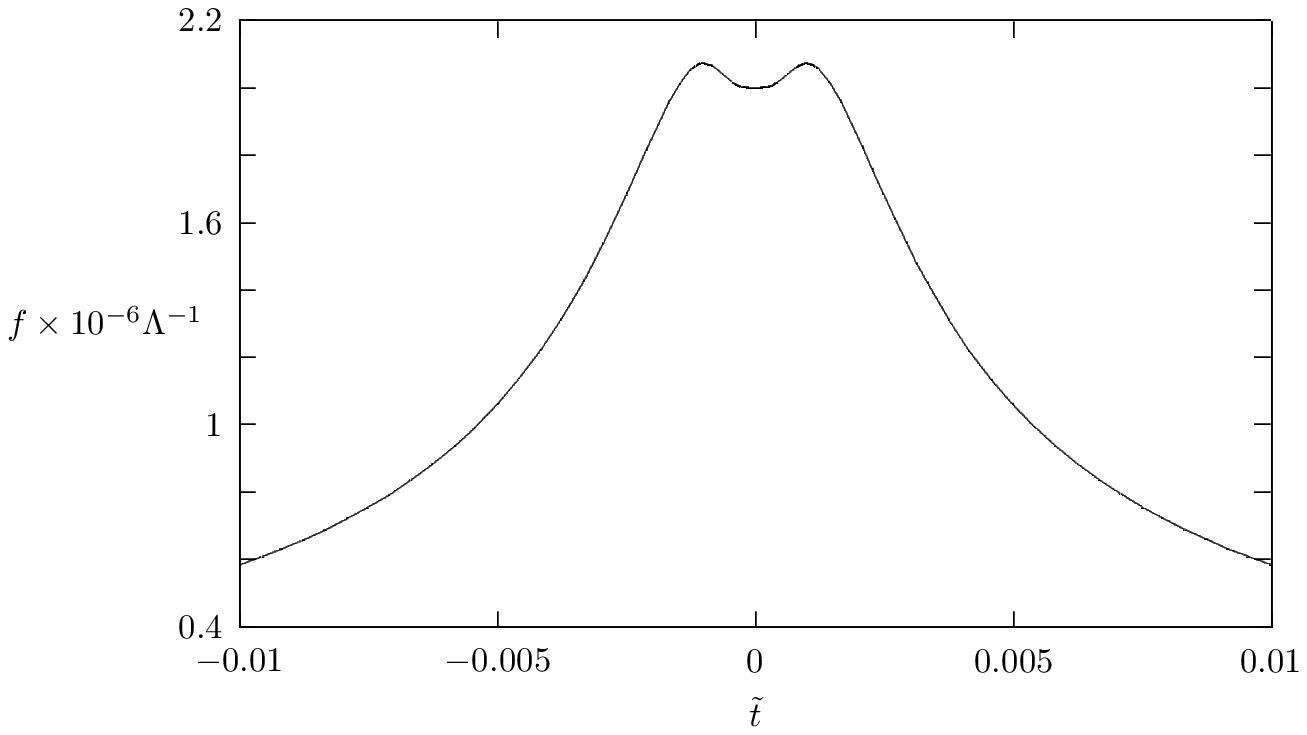}\\
\caption{Evolution of the function f in terms of the dimensionless
time $\tilde{t}$ in $\Lambda$CDM bounce cosmology.}
\end{figure}

\begin{figure}
\includegraphics{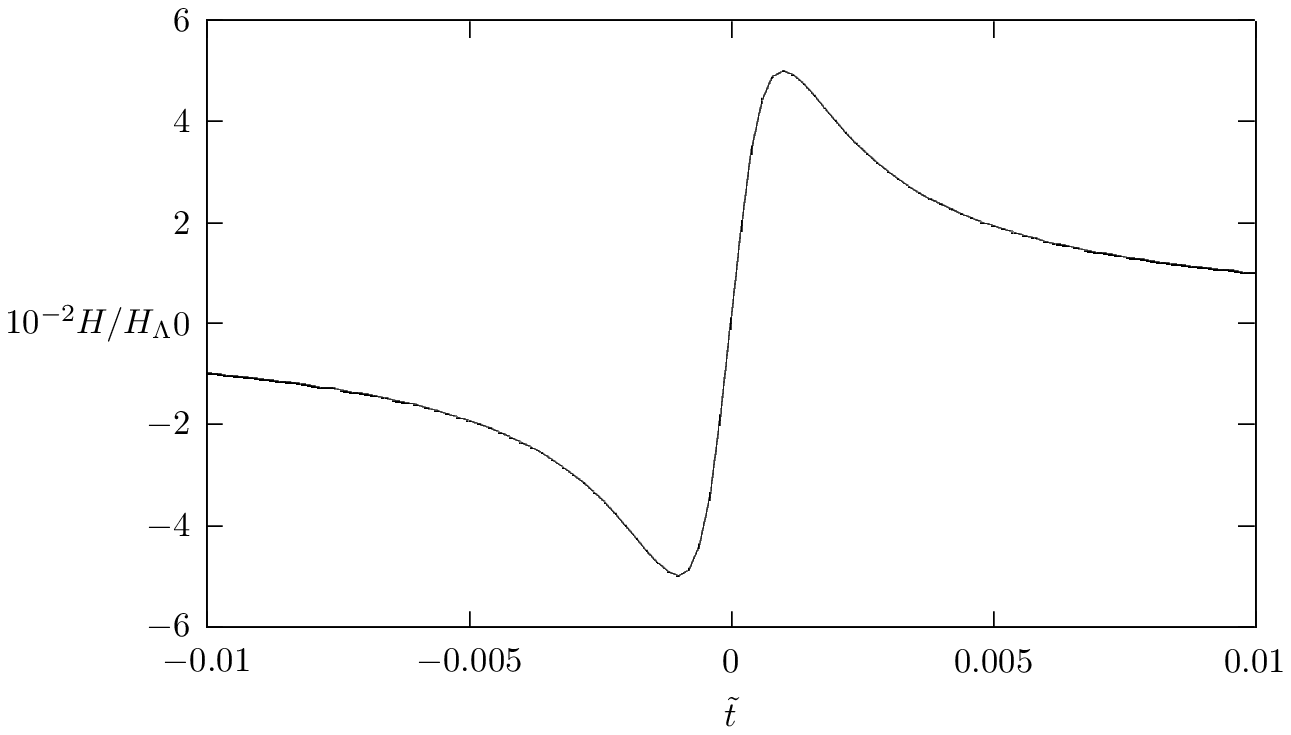}\\
\caption{Evolution of the Hubble parameter H in terms of the dimensionless
time $\tilde{t}$ in $\Lambda$CDM bounce cosmology.}
\end{figure}

The same method can be applied for constructing cosmological models without future singularities.
One consider two examples.

\textbf{Quasi-rip dynamics with bounce in future.} Let's consider the simplest case of phantom model with constant EoS parameter $w_{0}=p/\rho=-1-\epsilon/3=\mbox{const}$ ($g=\epsilon\rho/3$). In GR cosmology the evolution of scale factor as well known is
\be \label{BR}
a(t)=\frac{t_{f}^{2/\epsilon}}{(t_{f}-t)^{2/\epsilon}}.
\ee
We put $a(0)=1$. The moment of Big Rip singularity is $t_{f}=2\epsilon^{-1}\rho^{-1/2}_{0}$ ($\rho_{0}$ is the phantom energy density in the moment $t=0$). Let's choose the scale factor in form
\be \label{Bounce}
a_{B}(t)=\frac{(t_{f}^{2}+t_{1}^{2})^{1/\epsilon}}{(t_{1}^{2}+(t_{f}-t)^{2})^{1/\epsilon}},
\ee
where $t_{1}$ is the constant and $t_{1}<<t_{f}$. For $t<<t_{f}$ the $a_{B}(t)\sim a(t)$ but in the moment $t=t_{f}$ the bounce occurs instead Big Rip singularity. Phantom energy density is
\be\label{rhob}
\rho=\rho_{0}a^{\epsilon}_{B}(t)=\rho_{0}\frac{1+\delta^{2}}{\delta^{2}+(1-t/t_{f})^{2}},\quad \delta=t_{1}/t_{f}.
\ee
Substitution (\ref{Bounce}), (\ref{rhob}) into (\ref{EQFT}) gives the following expression for $f(t)$:
\be
f(t)=\frac{-2\rho_{0}}{\delta^2+(1-t/t_{f})^{2}}
\left\{(1+\delta^{2})+\frac{3(1-t/t_{f})^{2}}{\delta^{2}+(1-t/t_{f})^{2}}+2(1-t/t_{f})(1+\delta^2)
\frac{1}{\delta}\arctan\frac{1-t/t_{f}}{\delta}\right\}.
\ee
The function $f(t)$ is depicted on Fig. 4 for various $\delta$. In the moment of bounce $f(t)=-2\rho_{0}(1+\delta^{-2})$. On Fig. 5 the dependence $f$ from $T$ is presented. The effective EoS parameter is
\be
w_{eff}=-1-\frac{2\dot{H}}{3H^2}=-1-\frac{2\tilde{H}'}{3\tilde{H}^{2}},
\ee
where comma means differentiation on dimensionless time $\tilde{t}=t/t_{f}$ and $\tilde{H}=Ht_{f}$ is the dimensionless Hubble parameter:
\begin{equation}
\tilde{H}=\frac{2}{\epsilon}\frac{1-\tilde{t}}{\delta^2+(1-\tilde{t})^{2}}.
\end{equation}
The moment of ``dephantomization'' when $w_{eff}=-1$ is $\tilde{t}=1-\delta$. The moment of null acceleration ($w_{eff}=-1/3$) corresponds to $\tilde{t}=1-\delta\epsilon^{1/2}/(2+\epsilon)^{1/2}$.

\begin{figure}
\includegraphics{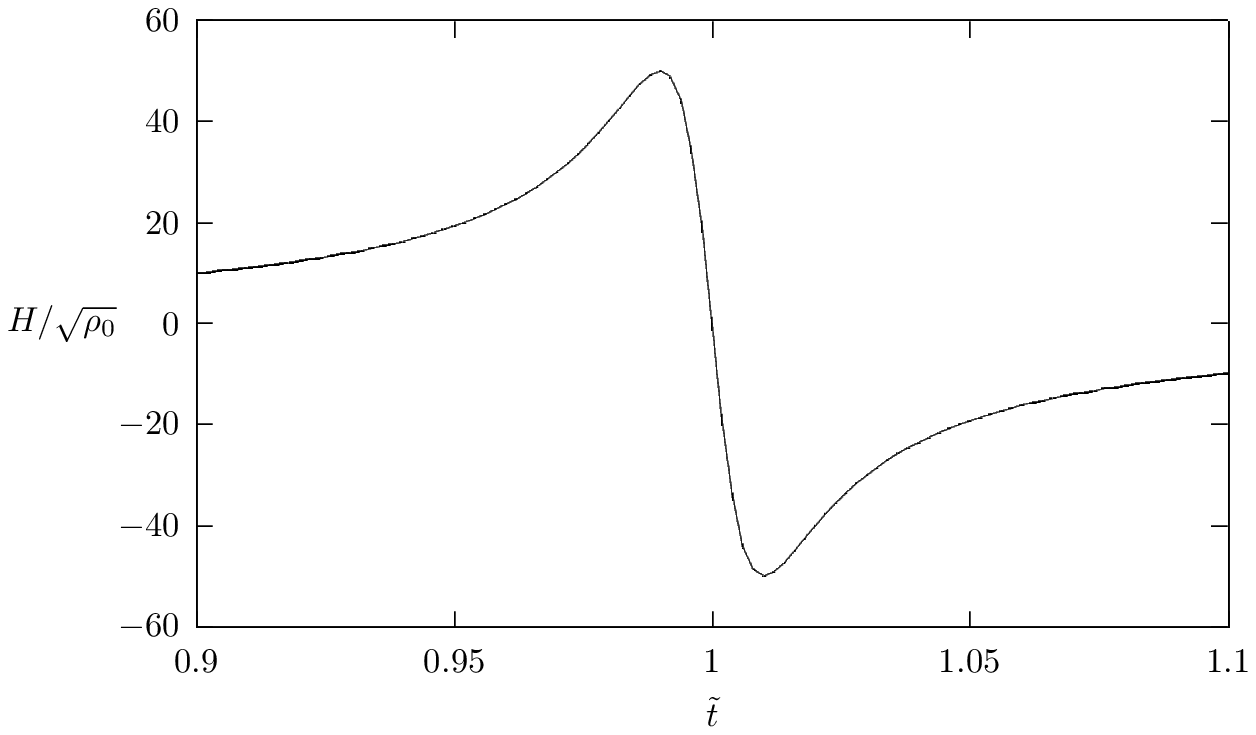}\\
\caption{Evolution of the Hubble parameter $H$ in terms of the dimensionless time $\tilde{t}$
in quasi-rip bounce cosmology.}
\end{figure}

\begin{figure}
\includegraphics{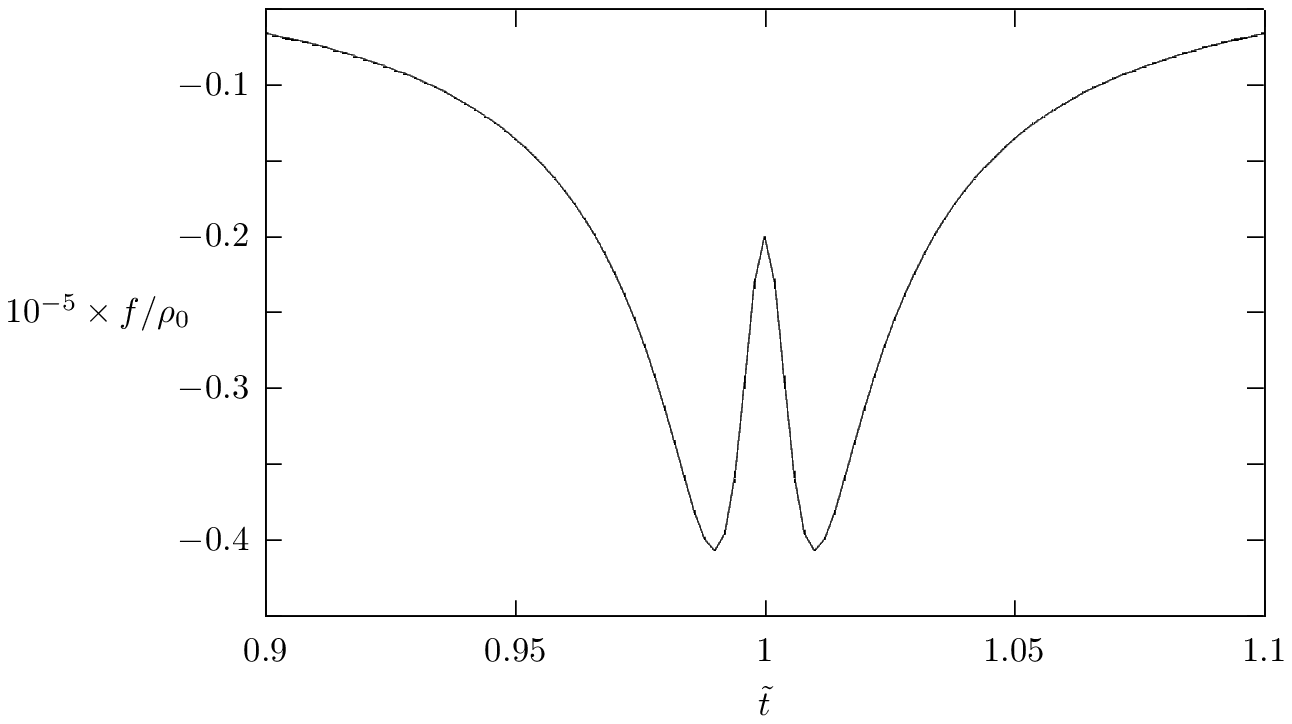}\\
\caption{Evolution of the function $f$ in terms of the dimensionless
time $\tilde{t}$ in quasi-rip bounce cosmology.}
\end{figure}

\textbf{Little rip dynamics with bounce in future.} The simple model with little rip (or sub-quantum potential -  $g(\rho)=4\alpha^{2}$) in GR cosmology leads to the following evolution:
\be\label{LRE}
a(t)=\exp(\alpha^2 t^{2}+\rho_{0} t),
\ee
where $\rho_{0}$ is energy density in the moment $t=0$. For energy density as function of scale factor we have
\be
\rho(a)=\rho_{0}+3\alpha^{2}\ln a.
\ee
One can consider the cyclic model (for simplicity we put $\rho_{0}=0$):
\be
a_{B}(t)=\exp\left(\alpha^{2}t_{b}^{2}\sin^{2}\frac{t}{t_{b}}\right).
\ee
For $t<<t_{b}$ the evolution of such universe coincides with Little Rip model in GR, but at the moment $t=\pi t_{f}/2$ the bounce occurs and universe begin contracts. The moment $t=\pi t_{f}$ corresponds to next bounce and universe expand again. The function $f(t)$ in interval $0<t<\pi t_{f}$
\be
f(t)=24\alpha^{4}t_{f}^{2}\sin\frac{t}{t_{f}}\left(\cos\frac{t}{t_{f}}-\sin\frac{t}{t_{f}}+\frac{2t}{t_{f}}\cos\frac{t}{t_{f}}\right)
\ee
It is obviously that $f(t+\pi t_{f})=f(t)$.

One note that described method of constructing non-singular solutions with bounce can be applied for brane $F(T)$ theory \cite{Bambam} and bounce loop-quantum cosmology from $F(T)$ gravity\cite{Amoros}. We also can consider the observational data \cite{Astashenok-4} for constraining model parameters. We plan investigate this issue in a near future.

\section{Conclusion}

In summary, various cosmological models in frames of $F(T)$ gravity are considered. The general scheme of constructing effective dark energy models with various evolution is presented. It is showed that these models in principle are compatible with $\Lambda$CDM model. The dynamics of universe governed by $F(T)$ gravity can mimics $\Lambda$CDM evolution in past but declines from it in a future. In addition we analyzed the some features which can take place in $F(T)$ gravity for universe filled real phantom energy with given EoS. It is showed that there are two possible mechanisms of avoiding final singularities in $F(T)$ gravity. The first is that in $F(T)$ gravity the rate of universe expansion can grows with energy density more slowly than in GR. Another possibility is bounce in future similar to that which may occur at the beginning of the universe.

\end{document}